\documentclass{kluwer}    
\usepackage{graphicx}

\begin{document}                                                              
                     
\begin{article}
\begin{opening}         
\title{Galactic Evolution along the Hubble Sequence} 
\author{Mercedes \surname{Moll\'{a}} \& Angeles I. \surname{D\'{\i}az}}  
\runningauthor{Mercedes Moll\'{a} \& Angeles I. D\'{\i}az}

\runningtitle{Galactic Evolution}
\institute{Departamento de F\'{\i}sica Te\'{o}rica, C-XI
Universidad Aut\'{o}noma de Madrid, Spain}
\date{September, 2002}
\begin{abstract}

A generalization of the multiphase chemical evolution model applied to
a wide set of theoretical galaxies is shown.  This set of
models has been computed by using the so-called Universal
Rotation Curve from \cite{per96} to calculate the radial mass
distributions of each theoretical galaxy. By assuming that the
molecular cloud and star formation efficiencies depend on the
morphological type of each galaxy, we construct a
bi-parametric grid of models whose results are valid in
principle for any spiral galaxy, of given maximum rotation
velocity or total mass, and morphological type.

\end{abstract}

\end{opening}           

\section{Introduction}

There exist three important correlations \cite{vil92,
oey93,zar94,dut99} among the observed characteristics of spiral and
irregular galaxies: 1) The mass-metallicity relation: the more
luminous the galaxy the higher its abundance; 2) The correlation
between abundance and morphological type: earlier type galaxies have
larger abundances than later type ones; and 3) The correlation between
the radial gradient of abundances and the morphological type: The
radial gradient of abundances is larger for later type galaxies than
for earlier ones.

The problem appears when we see that the mass-- metallicity relation
 seems to be the same for bright spirals and low-mass irregular
 galaxies while the correlation between the radial gradient and the
 morphological type has an effect on-off: when the galaxy mass
 decreases or the Hubble type changes to the latest irregulars, the
 steep radial gradient disappears and the abundance pattern becomes
 uniform for the whole disk.

To study correlations with a large number of galaxies for which we
need a large number of models.  For that, we build a bi-parametric grid
of theoretical models depending on the galaxy total mass and
morphological type.

\section{Ingredients of the multiphase chemical evolution model}

The scenario of the multiphase model begins with a spheroidal
protogalaxy sliced into cylindrical regions. The gas of the
protogalaxy has a mass M(R) computed from the rotation curve V(R),
which we take from \inlinecite{per96} to calculate radial mass distributions
M(R) for any value of the luminosity.

This gas collapses onto the equatorial plane and forms out the disk.
The infall rate is inversely proportional to a characteristic collapse
time scale, which changes according to the total mass of the galaxy:
\begin{equation}
\frac{\tau_{col,gal}}{\tau_{col,MWG}}=[\frac{M_{MWG}}{M_{gal}}]^{1/2}
\end{equation}

This implies that the characteristic collapse time scale
$\tau_{col,gal}$ is longer for the less massive galaxies.

There exist different phases of matter in each radial cylinder:
 diffuse gas, molecular clouds, massive stars, low and intermediate
 mass stars, and remnants. The star formation in the halo follows a
 Schmidt law. In the disk the star formation has 2 steps:
 molecular clouds are formed from the diffuse gas.  Then, stars form
 through cloud-cloud collisions, and from the interaction
 between massive stars and molecular clouds.

Every processes rates change along the galactocentric radius due
to the volume of each region through a proportionality coefficient
called efficiency or probability. Only the last described process does
not change with R being constant for all galaxies. Moreover, the
efficiency to form stars in the halo is assumed as constant for all
halos. Thus, only efficiencies to form molecular clouds and to form
 stars from collisions among these ones, are variable for each galaxy.  

The initial mass function was taken from \inlinecite{fer90} and the
nucleosynthesis prescriptions are taken from \inlinecite{woo95} for
$\rm M> 8 M_{\odot} $, \inlinecite{ren81} for stars $0.8 M_{\odot} < m
< 8 M_{\odot}$ and \inlinecite{iwa99} for SN-I.

\section{Model Results and Conclusion}

\begin{figure*}
\centerline{\includegraphics[width=12cm,height=6cm]{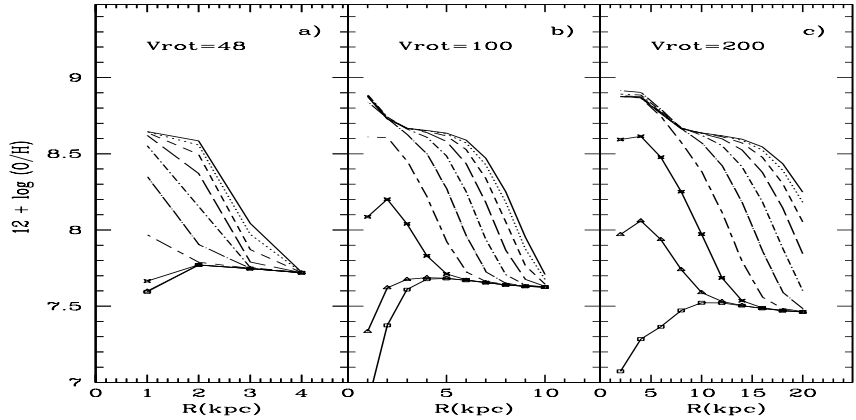}}
\caption[]{Radial of oxygen abundances for 3 different rotation
velocities (in km.s$^{-1}$) as marked in each panel, by assuming ten
different evolutionary rates or efficiencies for each one of
them. These ten values represent ten morphological types, from $T=1$,
the top solid line, up $T=10$, the bottom line with squares $\square$
as symbols.}
\label{oh}
\end{figure*}

We compute models for 44 radial mass distributions, and 10 different
values of efficiencies between 0 and 1 which are equivalent to 10
morphological types, for each one. More details about models and
results are described in \inlinecite{mol02}.

These models reproduce the 3 cited correlations as wee see in the
following two figures. In Fig.~\ref{oh} the radial distributions of
the oxygen abundances, given by $\rm 12 + log (O/H)$, are shown for 3
different rotation velocities as marked in panels.  Ten different
evolutionary rates or efficiencies, representing ten Hubble types,
from $T=1$, the top solid line, up $T=10$, the bottom line with
squares ($\square$) as symbols, are drawing for each one of them.

The more massive galaxies have strong abumndance radial gradients only for the
latest types ($T=6-9$), except $T=10$ which is flat, while the
earliest ones have very flat radial distributions.  The intermediate
mass galaxies show steep radial distributions for the intermediate
types ($T=4-8$), the latest ones being rather flat. The less massive
galaxies have no gradients for types later than 7, all the others
showing very steep radial distributions. The described behavior is in
agreement with the observed correlations. It is also clear from the
same Fig.~\ref{oh} that a minimum level of abundances is around 7.5
dex, which is the value found in the less enriched H{\sc II} regions.

\begin{figure}
\centerline{\includegraphics[width=12cm,height=6.2cm]{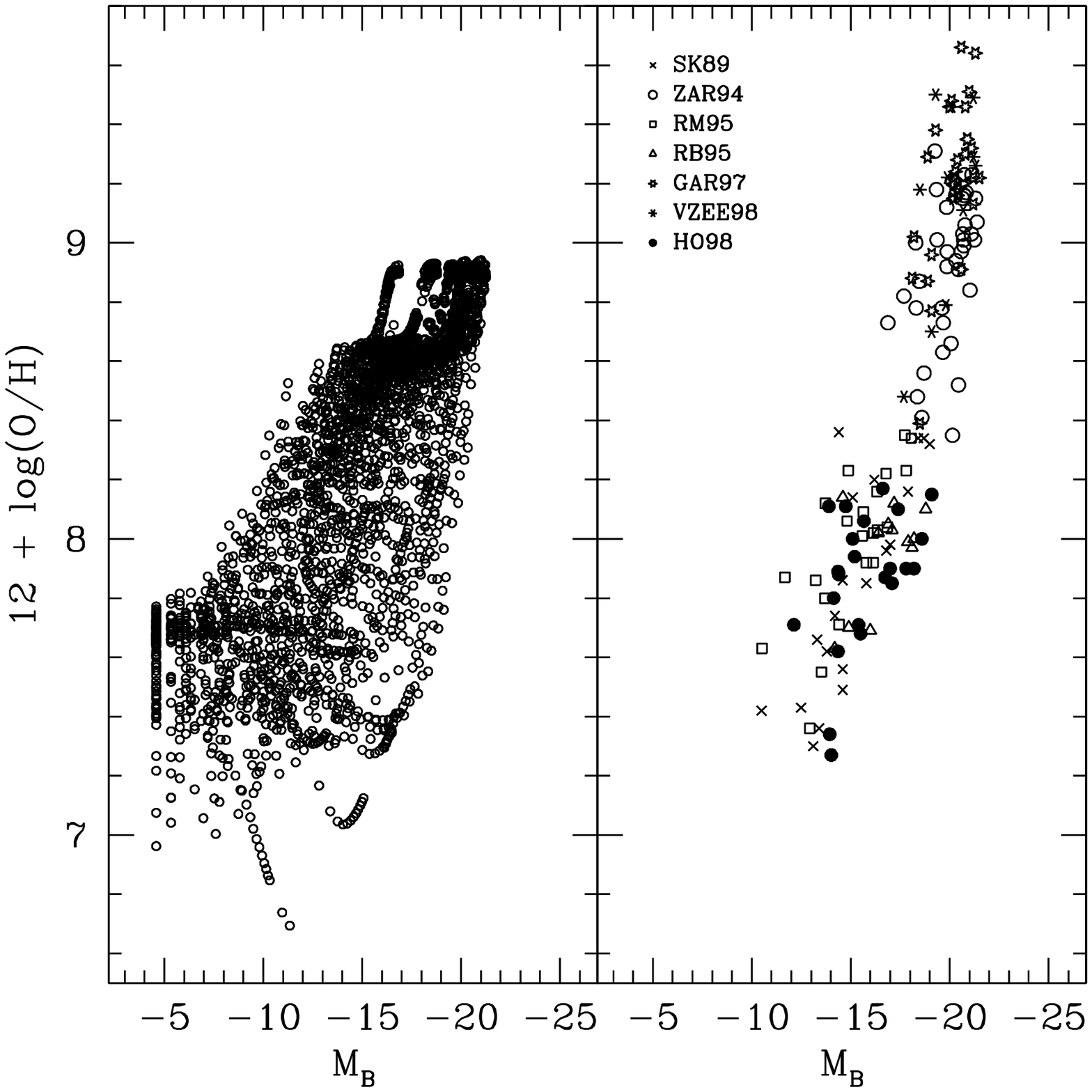}}
\caption[]{The mass-metallicity relation obtained for all radial regions
of our 440 models. a) The oxygen abundance $12 + log (O/H)$ {\sl vs} 
the stellar luminosity in each region and model
b) The data from  \inlinecite{ski89}--SK89--, \inlinecite{zar94}--ZAR94--,
 \inlinecite{ric95}--RM--, \inlinecite{ron95}--RB95--, \inlinecite{gar97}
--GAR97--, \inlinecite{vzee98} --VZEE98-- and \inlinecite{hid98} --HO98 
as marked in the panel}
\label{oh_mb}
\end{figure}

In Fig.~\ref{oh_mb}, panel a) we represent the same oxygen abundances
obtained with our 440 models for every radial regions as a function of
the stellar magnitude of the same region, computed by assuming a ratio
$\rm M_{*}/L=1$.  These relation can not be exactly compared with data
 --panel b)- which represent characteristic oxygen abundances {\sl vs} 
the magnitude of each galaxy, that is integrated quantities for a given 
galaxy, instead  the local characteristics represented by the models.
However, we can see that the correlation resulting from models have
the same behavior than observations.

Therefore, we conclude that our models follow adequately 
the observed trend for spiral and irregular galaxies.

\end{article}
\end{document}